\documentclass[11pt,twoside]{article}
\usepackage{cozumel2005}

\usepackage{epsf}
\usepackage{psfig}
\usepackage{lscape}
\def\mv{M_V}

\def\mk{\mbox{M}_{K}}
\def\mj{M_J}
\def\msol{M_\odot}

\def\mbol{M_{\rm bol}}
\def\mvol{{\rm M}_\odot{\rm pc}^{-3}}

\def\teff{T_{\rm eff}}

\def\mvol{\mbox{M}_\odot\, \mbox{pc}^{-3}}
\def\mlvol{(\log\mbox{M}_\odot)^{-1}\, \mbox{pc}^{-3}}

\def\simgr{\,\hbox{\hbox{$ > $}\kern -0.8em \lower 1.0ex\hbox{$\sim$}}\,}
\def\simle{\,\hbox{\hbox{$ < $}\kern -0.8em \lower 1.0ex\hbox{$\sim$}}\,}
\def\etal{{\it et al.\ }}

\begin{document}

\title{Review on low-mass stars and brown dwarfs }

\author{G. Chabrier$^1$, I. Baraffe$^1$, F. Allard$^1$ and P.H. Hauschildt$^2$}
\affil{$^1$Ecole Normale Sup\'erieure de Lyon,\\
Centre de Recherche Astrophysique de Lyon (UMR CNRS 5574),\\ 69364 Lyon
Cedex 07, France\\
$^2$Hamburger Sternwarte, Hamburg, Germany}

\begin{abstract}
In this review, we examine the successes and weaknesses of modern low-mass star and brown dwarf theory
from various comparisons with available experimental and observational constraints in different domains.
(1) We first focus on the mechanical (equation of state) and thermal (atmosphere) properties and on the
evolution of such cool and dense objects. We then examine the current shortcomings of the theory and
we discuss in detail recent observational analysis which have suggested discrepancies between the
models and the observations. (2) We then examine the stellar and brown dwarf
initial mass function and suggest that a power-law above the average thermal Jeans mass ($\sim 1\msol$)
rolling over a lognormal form below this limit adequately reproduces the observations of field and
young cluster stellar and brown dwarf distributions. This yields a reasonably accurate estimate of
the stellar and brown dwarf Galactic census. Finally (3) we examine the modern context of star
and brown dwarf formation and argue that the combination of turbulence driven fragmentation at
large scale and gravity at small scales provides an appealing solution
for the general star and brown dwarf formation mechanism. It also provides a physical ground for
the aforementioned power-law + lognormal form for the IMF, whereas a series of different power laws lacks such
a physical motivation. Finally we argue that the deuterium-burning limit as the distinction between stars
and planets has no physical foundation in this modern star formation scheme and should be abandoned.
Opacity limited fragmentation extending below 10 jupiter-masses down to a few jupiter masses, due to shocks, anisotropy
or magnetic fields, provides a much more robust limit, even though difficult to determine accurately.
Therefore, the various "direct" detections of exoplanets claimed recently in the literature are most likely regular low-mass brown dwarfs
and the direct detection of an extrasolar planet remains for now elusive.
\end{abstract}


\section{Introduction}
 
Ongoing observational projects
have revealed hundreds of substellar objects (SSO), brown dwarfs (BD) and gaseous planets orbiting stars outside the solar system.
These surveys
deliver large observational data bases and necessitate the best possible theoretical
foundation not only to understand the properties of these objects, including their formation mechanism,
 but also to evaluate properly their contributions to the baryonic content of the Galaxy.
Our group has ambitioned to derive a complete theory of
the structure, evolution and spectral signature of low-mass, dense
objects, from Sun-like to Saturn-like masses, covering 3 orders of magnitude in
mass and 9 in luminosity, bridging the gap between stars and gaseous planets.
We have incorporated the best possible physics aimed at describing the
mechanical and thermal properties of these objects - equation of state, synthetic spectra,
non-grey atmosphere models. We refer to previously published papers and reviews for a complete description of the
physics entering these models.
In this review we briefly examine the present status of this theory by confronting the theroretical predictions to various
experimental and observational results. We then explore the Galactic implications of the physics of low-mass objects, namely
mass function and Galactic census. Finally, we examine the formation of brown dwarfs and low-mass stars in the context of modern star formation theory.

\section{The physics of low-mass stars and brown dwarfs}

\subsection{Interior}

Central conditions for low-mass stars (LMS), i.e. objects below one solar mass, and brown dwarfs, i.e. objects below the hydrogen-burning minimum mass ($m_{HBMM}\simeq 0.07\,\msol$), range from  $T_c\sim 10^4$-$10^7$ K and
$\rho_c\sim 10$-$10^3$ g cm$^{-3}$, when spaning conditions from the Sun to Jupiter-mass interiors. Under these conditions, the average ion electrostatic
energy $(Ze)^2/a$, where $a=({3\over 4\pi}{V\over N_i})^{1/3}$ is the mean interionic distance,
is several times the average kinetic energy $kT$, characterizing a strongly coupled ion plasma.
The temperature is of the order of the electron Fermi temperature $kT_F$, so that the electron
gas is only partially degenerate. The temperature in the envelope is $kT<1$ Ryd, so electronic
and atomic recombinations take place in this region. Finally, the electron binding energy
is of the order of the Fermi energy $Ze^2/a_0\sim E_F$ so that {\it pressure} dissociation
and ionization take place along the interior profile. Within the past recent years, several high-pressure
shock wave experiments have been conducted in order to probe the equation of state (EOS) of deuterium,
the isotope of hydrogen, under conditions characteristic of the interior of these objects. Gas gun shock compression experiments were
generally limited to pressures below 1 Mbar \cite{Nellis83}, probing only the domain
of molecular hydrogen. The Saumon-Chabrier-VanHorn EOS (Saumon, Chabrier \& VanHorn 1995) was found to adequately reproduce the experimental pressure-density profile but
to yield temperatures about 30\% higher than the gas gun results, indicating insuffisant H$_2$
dissociation. A slightly revised version recovers well these gas gun experimental results \cite{Saumonetal00}.
Modern techniques include laser-driven shock-wave experiments \cite{Collinsetal98}, pulse-power compression experiments \cite{Knudsonetal04} and
convergent spherical shock wave experiments \cite{Belovetal02, Boriskovetal03} and
 can achieve higher pressures than gas gun
experiments, namely up to 5 Mbar on liquid D$_2$ at high temperature, exploring
for the first time the regime of pressure-dissociation and ionization. Unfortunately, these recent experiments
give different results at high pressure and this controversy remains to be settled before robust comparison between
experiment and theory can be made in
 the very domain of hydrogen pressure-ionization, i.e. $P\sim 1$-3 Mbar. For pressures above a few Mbar, hydrogen and helium become fully ionized and accurate Monte Carlo simulations of the
 thermodynamic properties can be used as a benchmark for the EOS. These high-pressure experiments clearly open a new window in physics and astrophysics by
constraining the EOS of stellar, brown dwarf and planet interiors in laboratory experiments.

\subsection{Atmosphere}

Atmosphere equilibrium condition $d\tau = \bar \kappa dP/g$, where $g=Gm/R^2\approx 10^3-10^5$
cm s$^{-2}$ is the characteristic surface gravity of SSOs, yields $P_{ph}\sim g/\bar \kappa
\approx 0.01-1$ bar and $\rho_{ph}\approx$10$^{-6}$-10$^{-5}$ g cm$^{-3}$ near the photosphere. Collision effects are important under such conditions and induce new sources of absorption
like e.g. the collision-induced absorption between H$_2$-H$_2$ or H$_2$-He below $\sim 5000$ K \cite{Linsky69, Borysowetal}.
For effective temperatures characteristic of LMS and BDs
($\teff \simle 5000$ K), numerous molecules
form, in particular metal oxydes and hydrides (TiO, VO, FeH, CaH, MgH), the major absorbers in the optical,
and CO, H$_2$O which dominate in the infrared.
The situation becomes even more complex for temperatures below $T\sim 2500$ K, i.e. for atmospheric
conditions of objects ranging from the coolest stars to BDs and jovian planets.
For $\teff \simle 2500$ K,
i.e. at the bottom of the main sequence, there is evidence for condensation of metals and silicates into grains (e.g. TiO into CaTiO$_3$, Mg, Si into
MgSiO$_3$, Ca into CaSiO$_3$, Ca$_2$SiO$_4$)  {\cite{Tsujietal96, JonesTsuji97, Leggettetal98, Lodders99, BurrowsSharp99, Burrowsetal00, Allardetal01}.
At 2000 K, most  of
the carbon is locked into carbon monoxide CO, while the oxygen is found  in
titanium TiO and vanadium VO monoxides (dominating the optical absorption) and
water vapor H$_2$O (shaping the infrared spectrum). This absence of refractory metal line absorption signature
in the spectrum defines the so-called "L-dwarf" domain.
Below $\sim 1800$ K, the dominant equilibrium form of carbon is no longer CO but CH$_4$ \cite{FegleyLodders96},
as identified for e.g. in Gl229B \cite{Oppenheimeretal95, Allardetal96, Marleyetal96}.
As confirmed by the observation of Gl229B {\cite{Oppenheimeretal95},
methane bands begin to appear in the infrared (1.6 and 2.2 $\mu$m), while
titanium dioxide and silicate clouds form at the expense of TiO, modifying
profoundly the thermal opacity of the atmosphere. The presence of methane absorption in the spectrum defines the "T-dwarf" spectral type, even though weak CH$_4$
absorption is already present in the latest L dwarfs.
For jovian-like atmospheres, the dominant equilibrium form of
nitrogen
N$_2$ is NH$_3$ ($\teff \simle 600$ K) and below $\teff \sim 600$ K (depending on gravity) water
condenses to clouds at or above the photosphere. 
Figures \ref{FigMLT} displays the spectral evolution of M, L and T dwarfs, with identified main absorption sources.
In spite of all these complex phenomena, BDs radiate nearly 90\% of their
energy at wavelengths longward of $1 \, \mu$m, with a peak around $\sim$ 1-2 $\mu$m,
and infrared
broad-band colors are prefered to optical ones for the identification of these objects.

\begin{figure}[!ht]
\plotone{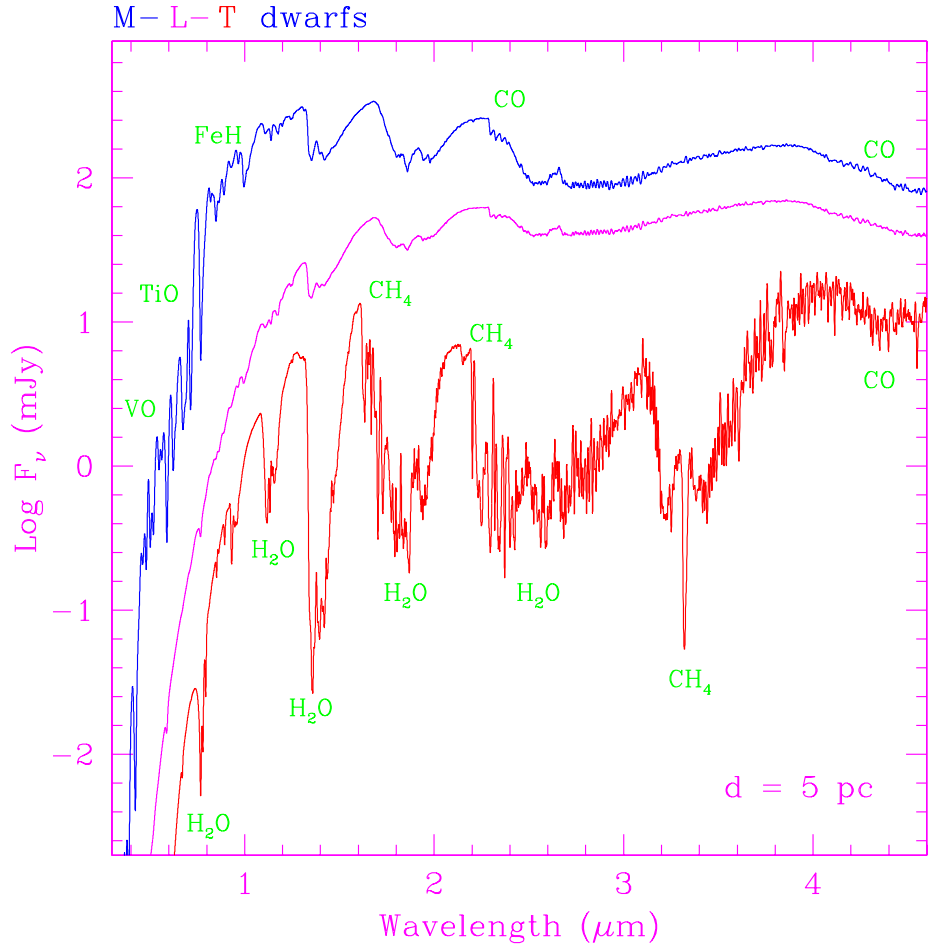}
%\vspace{4.8in}
\caption{Spectral energy distributions of typical M, L and T dwarfs according to models of Allard et al. (2001).
From top to botton, $\teff=2500$, 1800 and 1000 K, respectively, with surface gravities and radii from Chabrier et al. (2000) for an age 5 Gyr.
The most important bands of absorption are indicated.}
\label{FigMLT}
\end{figure}

The condensates or grains affect the atmosphere in different ways. Grain formation depletes the corresponding gas-phase absorber in certain regions of the
atmosphere and modifies the EOS itself
and thus the atmospheric temperature profile, but it also strongly modifies the opacities
and thus the emergent spectrum. At last it produces an increase of the temperature in the
uppermost layers of the atmosphere, where lines form, the backwarming effect, so that molecular bands like e.g. H$_2$O form in hotter regions and are thus weaker. This produces a significant reddening of the colors in late M and L dwarfs. In the opposite, the strong absorption
bands of methane in the infrared (H, K and L bands) induce a redistribution of the emergent flux at shorter wavelengths,
yielding bluer colors.
Our previous calculations of main sequence objects {\cite{BCAH98}, based on the atmosphere models of Hauschildt et al.
(1999) have been extended by taking into account this grain formation process in two extreme regimes. In the so-called "dusty" limit, all condensed species are included explicitely both in the atmosphere EOS and in the radiative
transfer equation \cite{Allardetal01} and provide the proper boundary conditions for consistent evolutionary calculations for such "dusty" objects \cite{Chabrieretal00a} in the effective temperature range $1700\la \teff \la 2500$ K.
At low enough temperature, $\teff \la 1700$ K, we consider the other so-called "condensed" limit, where all the grains either form
or have sunk below the photosphere, halting the redward progression of infrared colors \cite{Baraffeetal03}. Several objects have now been discovered in the
in-between "early T-dwarfs" domain, with $0.5\la J$-$K\la 1.5$ \cite{Kirkpatricketal00, Burgasseretal02}. In this regime,
grain dynamics with competing convective and settling time scales must be considered \cite{AckermanMarley01, Allardetal03}, yielding complex, and possibly non-monotonic color-distributions
with brightening J luminosities for the earliest T dwarfs \cite{TinneyBurgasserKirkpatrick03, Knappetal04, Vrbaetal04}.

The "dusty" models have been shown to reproduce the spectra and colors of various early L-dwarfs including the
ones discovered by the DENIS project \cite{Ruizetal97, Tinneyetal98} and GD165B, which lies on the very edge of the H-burning limit {\cite{Kirkpatricketal99} while
the "condensed' models give a good representation of the T-dwarf spectral energy distribution and photometry \cite{Allardetal96, Baraffeetal03}.
In the domain between late L-dwarfs and T-dwarfs, as mentioned earlier, models including dynamical processes must be considered. Atmosphere models
including cloud
sedimentation seem to give at least a qualitative description of the L to T transition \cite{Marleyetal02, Allardetal03}. No consistent evolutionary
calculation, however,
exists in this domain yet. Fortunately, this late-L early-T dwarf domain concerns only massive and/or young BDs, and thus
a limited fraction of the global BD population.

Although the general description of BD spectral evolution can be considered as satisfactory,
there is definitely
room for improvement in a few bands, where the observed flux is either overestimated or underestimated by the models.
Such discrepancies are mostly tied to incomplete opacities (e.g. TiO, H$_2$O, CH$_4$, CaH, VO, FeH).
An important issue in the T-dwarf domain concerns the alkali-line pressure broadening like e.g.
the Na I (0.59 $\mu$m) and K I (0.77 $\mu$m) resonance doublets \cite{Tsujietal99, Burrowsetal00, BurrowsX, AllardAllard} which absorb the flux in the optical,
yieding redder optical colors with decreasing $\teff$, and accurate treatments of the far wings of these lines are necessary to model accurately very cool BDs.
The spectral evolution of BDs is rendered even more complicated by the occurence of non equilibrium thermochemical reactions, as observed e.g. by the greater absorption
of CO and thus an enhanced
abundance than predicted by chemical equilibrium models \cite{NollGeballeMarley97, Saumonetal03}. At last,
there is strong observational suggestion that the cloud surface coverage breaks up in the L-T transition region, as suggested by the aforementioned increasing J flux, which could
then emerge from deeper layers of the atmosphere \cite{Burgasseretal02}. Spectra changes in this domain, i.e. around $\sim 1300$ K, are thus not due
to decreasing temperature, this latter remaining constant over the $\sim$ L7-T4 domain, but could be explained by clearing of the cloud decks \cite{Burgasseretal02, Tsuji05}.

\section{Evolution}

\subsection{ Color-magnitude diagrams of disk and young cluster objects}

\begin{figure}[!ht]
\plotone{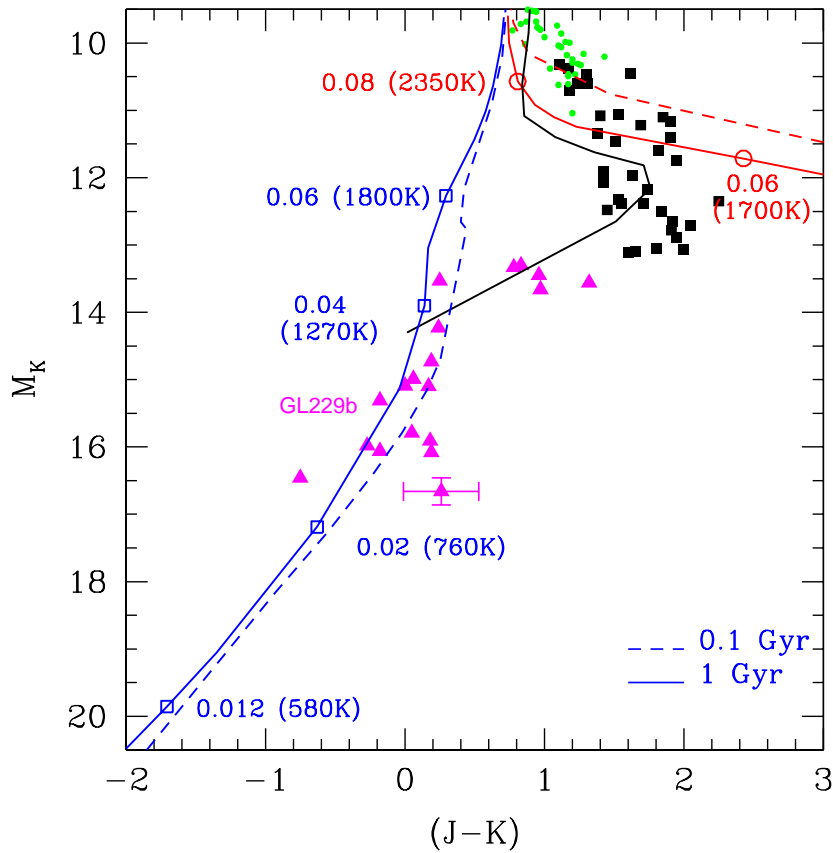}
%\vspace{2.8in}
\caption{Near-IR $K-(J-K)$ color-magnitude diagram.
 The data are from Leggett (1992) (dots), Dahn et al. (2000) (squares) and Reid et al. (1999) (filled triangles).
Theoretical models correspond to various atmosphere models and ages:
DUSTY (right) and COND (left) models. Preliminary models for late L early T are also shown between these two limits.
Masses (in $\msol$) and $\teff$
are indicated for the DUSTY and COND isochrones.}
\label{Figj-k}
\end{figure}

As mentioned above, present evolutionary calculations have been conducted with three different atmosphere models: (i) for objects hot enough, i.e.
massive or young enough, to
preclude the formation of grains ($\teff \simgr 2800$ K), we have used the grainless so-called
"NextGen" atmosphere models \cite{Hauschildtetal99, BCAH98}; (ii) for objects below 2800 K, i.e. the L-dwarf regime, which encompasses
the bottom of the main sequence, we have used atmosphere models
wich include the grain opacity sources in the transfer equation \cite{Allardetal01},
the so-called "dusty" models {\cite{Chabrieretal00a};
(iii) for objects below about 1500 K, down to Jupiter-like temperatures, we have considered
cases where the condensates settle rapidly below the photosphere and - although modifying
the atmosphere EOS - do not participate to the opacity, the aforementioned "condensed" models \cite{Baraffeetal03}.
These latter calculations are
motivated by the absence of grain features in the atmosphere of objects below $\teff \sim 1000$ K,
i.e. Gliese 229B-like objects.

Figure \ref{Figj-k} displays a K vs J-K color-magnitude diagram (CMD).
In terms of colors there is a competing effect between grain and molecular sources of absorption for objects at the bottom of and just below the main sequence,
with grain opacity
yielding eventually a severe redshift of the colors, a consequence of the aforementioned backwarming effect.
For cooler objects, CH$_4$ absorption in the infrared leads to a blueshift
for infrared colors. The figure also displays preliminary calculations based on atmosphere models including grain dynamics \cite{Allardetal03} in the late-L early-T domain.

Several BD surveys have been conducted in young clusters and it is important to develop
accurate non-grey models for PMS stars and young BDs. The more massive of such
objects are too hot for grain formation to occur. Although the Baraffe et al. (1998) models give a consistent description of the CMDs
of these clusters \cite{Luhmanetal03}, the situation for young objects is not as satisfactory as for objects with higher gravity.
As discussed in detail in Baraffe et al. (2002) both models and observations are hampered by numerous uncertainties and great caution must be taken
when considering young age ($\la 10$ Myr) objects.
Note that, when using the Luhman (1999) $\teff$-spectral type scale for young objects adjusted to yield consistent,
coeval sequences for the quadruple system GG-tau, whose component masses extend from 1.2 $\msol$ down to $\sim 0.035\,\msol$ \cite{Whiteetal99},
these LMS and BD models give {\it coeval} sequences for several other clusters like e.g. IC348, Taurus, Chamaeleon \cite{Luhmanetal03}.
Although the possibility that such an agreement be fortuitous can not be totally excluded, the fact that these sequences are composed of hundred
of sources lends confidence to the reliability of the models.

\subsection{Mass-magnitude relations}

Combining adaptative optics and accurate radial velocity surveys, Delfosse et al. (2000)
 and Segransan et al. (2003) have been able to determine mass-magnitude relationships (MMR) of nearby low-mass binaries down to the vicinity of the H-burning limit.
Figure \ref{FigmassMv} displays the comparison of the Andersen (1991) and Segransan et al. (2003) data in the V band with different theoretical MMRs, namely the parametrizations of Kroupa et al. (1993) (KTG), Reid et al. (2002) for $\mv <9$
complemented by Delfosse et al. (2000) above this limit and the models of Baraffe et al. (1998) (BCAH98) for two isochrones.
The KTG MMR gives an excellent parametrization of the data
over the entire sample but fails to reproduce the flattening of the MMR near the low-mass end, which arises from the onset of degeneracy near the bottom of the main sequence \cite{CB00}, and thus yields a too steep slope.
The Delfosse et al. (2000) parametrization, by construction, reproduces the data in the $\mv$=9-17 range. For $\mv< 9$,
however, the parametrization of Reid et al. (2002) misses a few data, but more importantly does not yield the
correct magnitude of the Sun for its age. The BCAH98 models give an excellent representation
for $m\ga 0.4\msol$. Age effects due to stellar evolution start playing a role above $m\sim 0.8\,\msol$,
where the bulk of the data is best reproduced for an age 1 Gyr, which is consistent with a
stellar population belonging to the young disk ($h< 100$ pc).
Below $m\sim 0.4 \,\msol$, the BCAH98 MMR clearly differs from the Delfosse et al. (2000) one. Since we
know that the BCAH models overestimate the flux in the V-band, due to still incomplete molecular opacities \cite{Chabrieretal00a},
mass function calculations (\S4) have been conducted with the Delfosse et al. (2000) parametrization in this domain.
The difference yields a maximum $\sim 16$\% discrepancy in the mass determination near $\mv \sim 13$ \cite{Chabrier02}.
Figure \ref{FigmassMag} displays a comparison of the data with the BCAH98 calculations in all available
observational bands. As seen in the figure, there is
a remarkable agreement between the theoretical {\it predictions} and the data in all available infrared bands, the most favorable domain for LMS and BD observations.
Overall, the general agreement can be considered as excellent, and the error
on mass functions derived when using these MMRs is smaller than the observational error bars in the LF.

\begin{figure}[!ht]
%\plottwo{Plot_various_logmMv_2.ps}{Plot_massMag_Segransan.ps}
\plotone{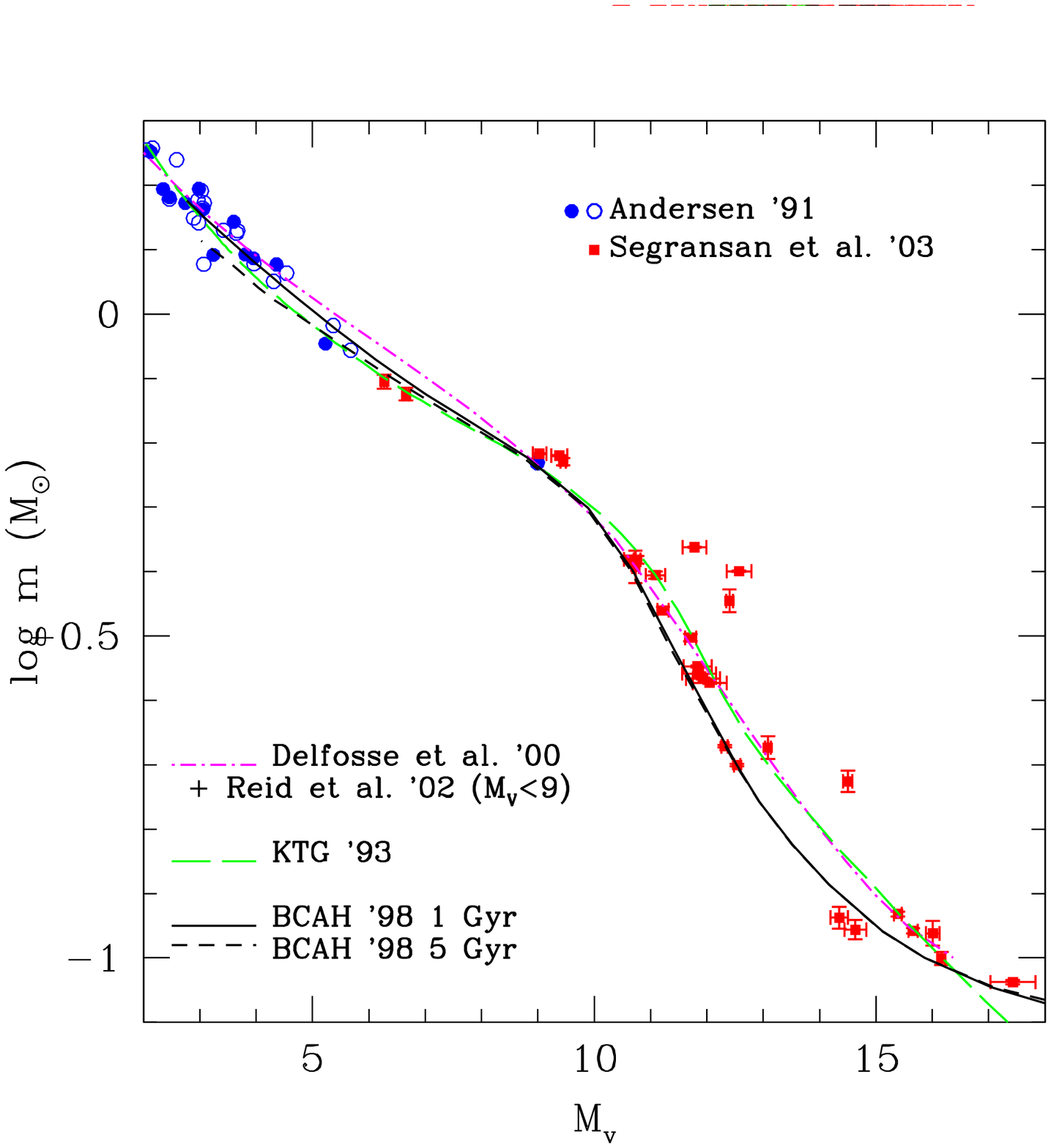}
\caption{Comparison of various m-$\mv$ relations}
\label{FigmassMv}
\end{figure}
\begin{figure}
\plotone{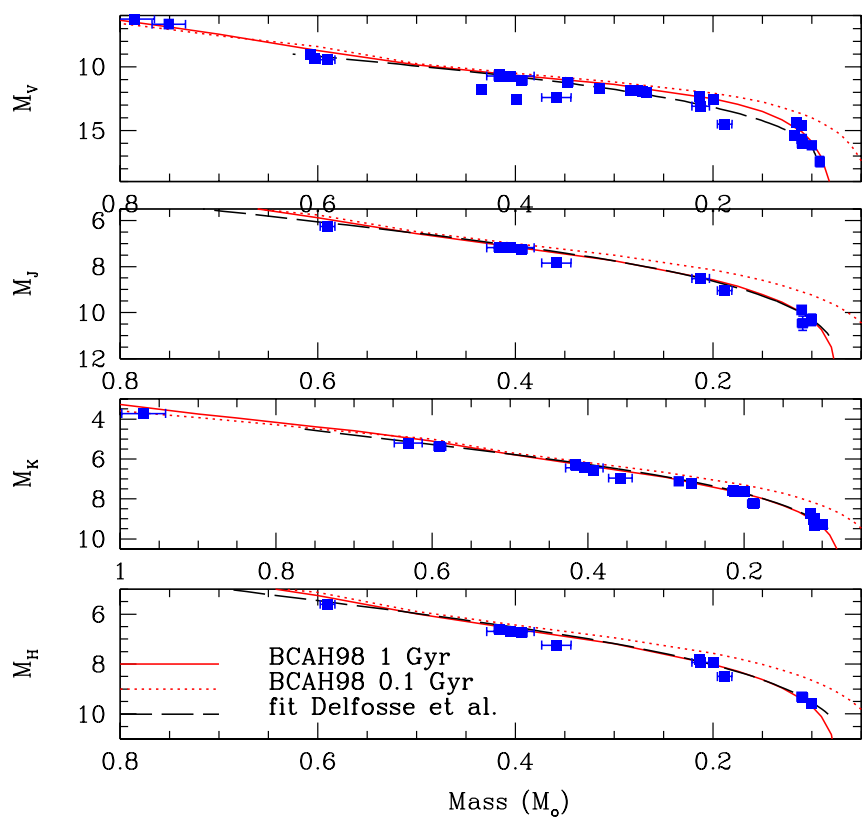}
\caption{Comparison of oberved \cite{Seg03b} and theoretical \cite{BCAH98} mass-magnitude relationships in various bands}
\label{FigmassMag}
\end{figure}

\begin{figure}[!ht]
\plotone{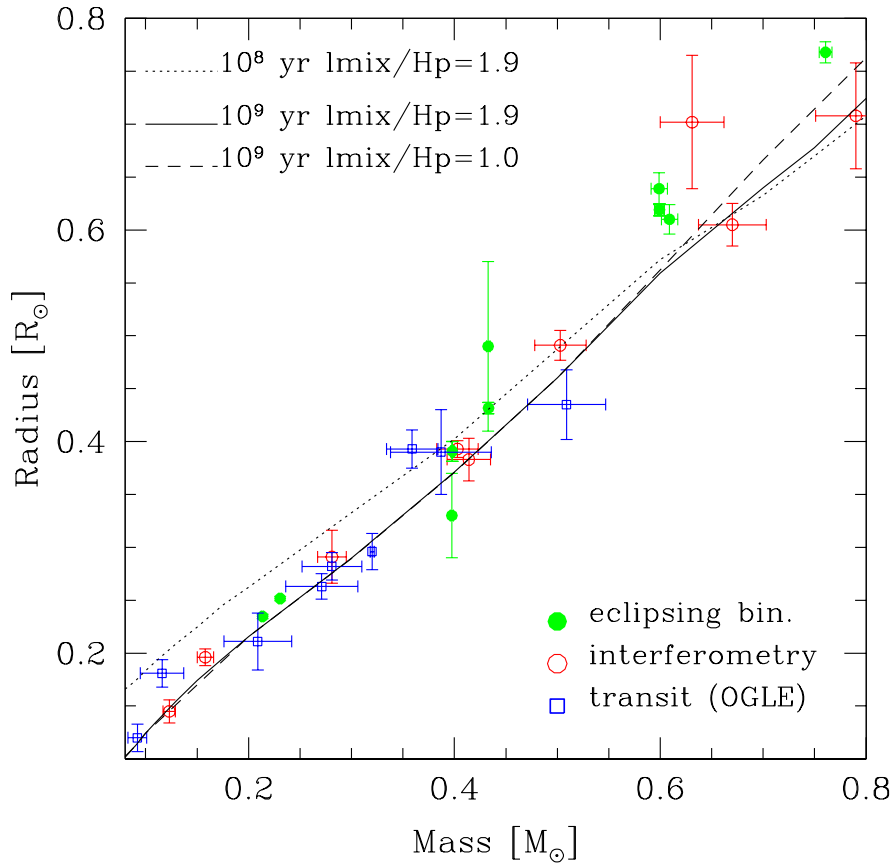}
%\vspace{2.8in}
\caption{Comparison of the theoretical (Chabrier \& Baraffe 2000) mass-radius relationship for low-mass stars with observations}
\label{FigMassrad}
\end{figure}

\subsection{Mass-radius relation}

The radii of many LMS have been determined accurately from various techniques. Eclipsing binaries of course provide the most natural method but include only
a limited number of systems below 1 $\msol$. Interferometry with the VLTI allows a precise determination of the radii of nearby binaries \cite{Seg03},
while transit observations from the OGLE microlensing survey improve significantly the statistics. Figure \ref{FigMassrad} shows the comparison between the theoretical and observed
radii from 0.8 $\msol$ down to the HBMM, including the determination of the radius of the presently smallest H-burning object \cite{Pontetal05}. The remarkable agreement
between theory and observation gives confidence in the underlying physics used to determine the mechanical structure of these
cool and dense objects, in particular the EOS (see \S2). The observed radii for the eclipsing binaries, however, are found to be $\sim 10$\% larger than the theoretical ones \cite{TorresRibas02}.
As seen on the figure, however, all eclipsing
binary radii but one lie above the other observed radii. It must be noted that all these eclipsing binaries are magnetically very active, a consequence of dynamo
generated processes in objects at small orbital separation. It is thus natural to imagine that spot area cover a significant fraction of the irradiating surface
of these objects, yielding smaller contraction along evolution.
Although purely speculative at
the present stage, this suggestion provides a viable explanation for the larger radius of eclipsing binaries compared to other binaries.

\section{Stellar and brown dwarf mass functions}

\subsection{Galactic field}

A proper census of the number of stars and BDs bears important consequences not only for
an accurate determination of the Galactic mass budget (about 70\% of the Galactic mass is under the form of objects below a solar mass) but for our understanding of star formation.
The determination of the stellar IMF can be done easily from observed luminosity functions (LF), using
accurate mass-magnitude relationships (MMR). Using the MMRs mentioned in \S3.2, or fits
based on observed nearby binaries \cite{Del00}, the stellar IMF has been determined recently
by Chabrier (2003, 2005), based on observed V-band and K-band LFs. Such an IMF, and comparison with other IMFs, is portrayed in Figure \ref{FigIMF}.
The solid line superposed to the IMFs derived from the LFs illustrates the following analytical parametrization (in $\mlvol$):

\begin{eqnarray}
\xi(\log \,m)&=&0.093\times \exp\Bigl\{-{(\log \, m\,\,-\,\,\log \, 0.2)^2\over 2\times (0.55)^2}\Bigr\},\,\,\, m\le 1\,\msol \nonumber \\
&=&0.041\,m^{-1.35\pm 0.3}\,\,\,\,\,\,\,\,\,\,\,\,\,\,\,\,\,\,\,\,\,\,\,\,\,\,\,\,\,\,\,\,\,\,\,\,\,\,\,\,\,\,\,\,\,\,\,\,\,\,\,\, , m\ge 1\,\msol
\label{IMF1}
\end{eqnarray}

This IMF differs slightly from the one derived in Chabrier (2003), which was based on the 5-pc LF \cite{Dahnetal86, HmcC90},
whereas the present one is based on the revised 8-pc LF \cite{Reid02}.
The difference at the low-mass end between the two parametrizations reflects the present uncertainty at the faint end of the disk LF, near the H-burning limit (spectral types $\ga$ M5).
Clearly a better determination of the faint part of the disk LF is required
before the IMF can be determined with higher accuracy at the H-burning limit.
Note that the field IMF is also
representative of the bulge IMF (triangles), derived from the LF of Zoccali et al. (2000).

The determination of the BD IMF is a complicated task. By definition BDs never reach
thermal equilibrium and most of the BDs formed at the early stages of the Galaxy
have now fainted to very faint luminosities, below present limits of detection.
Observations are thus biased towards young and/or massive BDs. Monte-Carlo calculations, taking into account a mass probability distribution function $P(m)$, given by the mass function, and a
{\it time} probability distribution function, $P(t)$, given by the star formation rate (SFR) must be conducted to obtain the BD IMF (assuming the mass- and age-distributions can be separated out). Such
calculations have been carried out by Reid et al. (1999), Chabrier (2002, 2003, 2005), Burgasser (2004) and Allen et al. (2005).
These calculations can be confronted to present observations of field BDs.

\begin{figure}[!ht]
\plotone{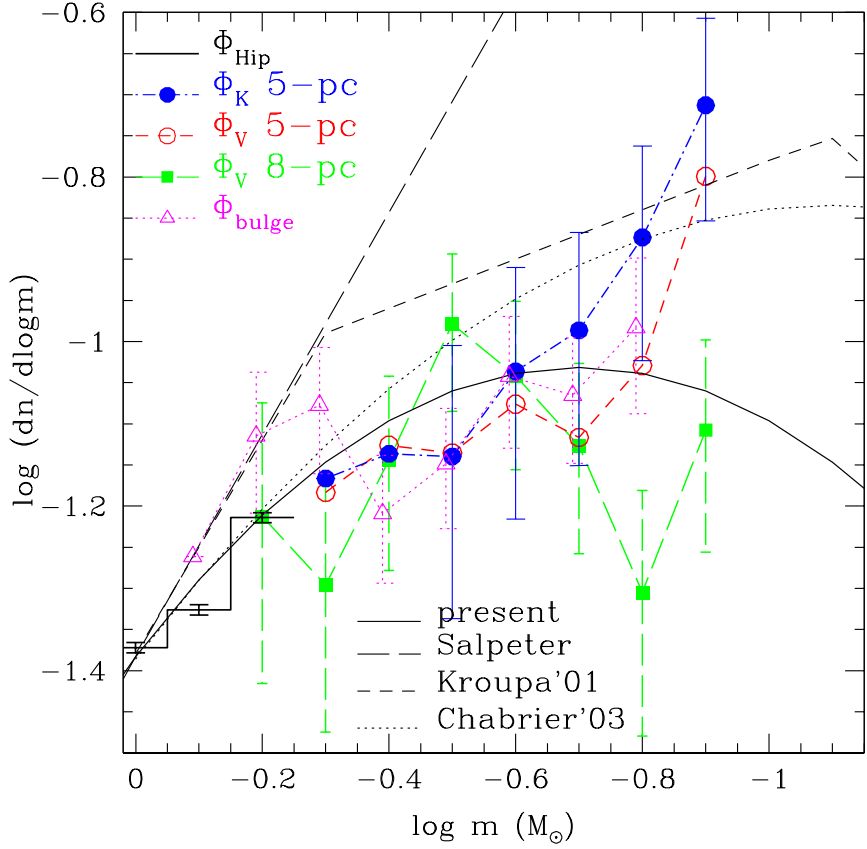}
%\vspace{2.8in}
\caption{Disk IMF determined from various luminosity functions $\Phi$}
\label{FigIMF}
\end{figure}

\begin{figure}[!ht]
\plotone{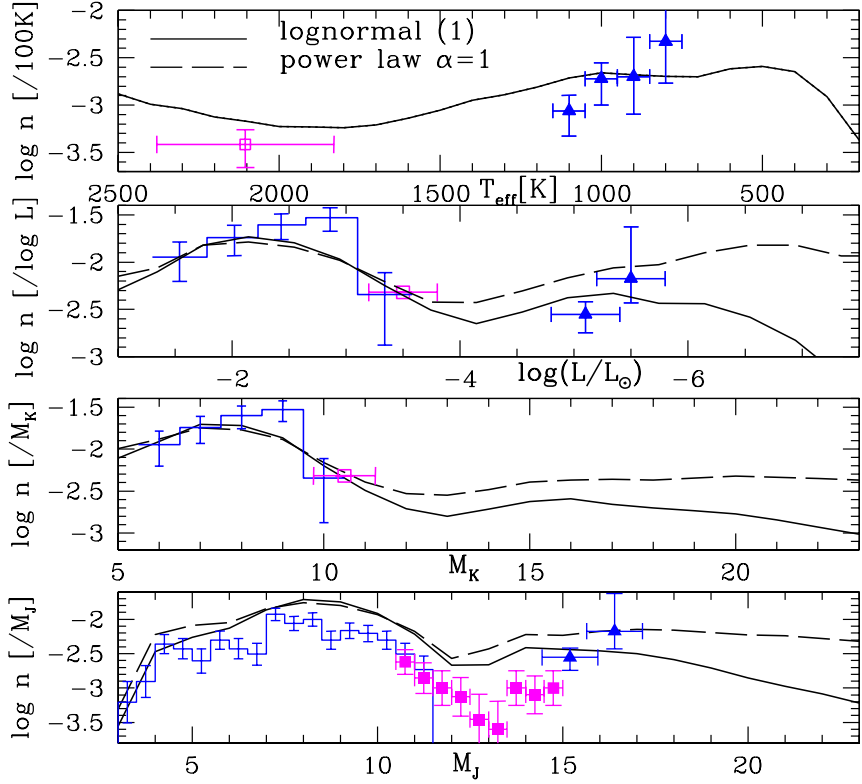}
%\vspace{2.8in}
\caption{Comparison of the observed LMS and BD distributions with the ones obtained with the IMF (1)
(solid line) and with a power law IMF ${dn\over dm}\propto m^{-1}$, with the same normalization at 0.1 $\msol$ (dash-line).}
\label{FigBDLF}
\end{figure}

Figure \ref{FigBDLF}
displays the calculated BD density distributions as a function of $\teff$, $L$, $\mk$ and $\mj$,
obtained by aforementioned Monte Carlo calculations, using the BD cooling models developed in the Lyon group, and the most recent estimated LMS and BD densities \cite{Gizis00, Burgasser01,Cruz04}.
These calculations include the correction for unresolved binaries. Binary fractions are based on the observed values
i.e. a binary frequency decreasing from $\sim 50\%$ to $\sim 20\%$ with decreasing mass from G-type stars to BDs.
The agreement between the theoretical calculations and the observations is rather satisfactory, keeping in mind the remaining uncertainties
both in BD cooling theory (in particular in the late L early T transition domain) and in the observational determinations of the BD $\teff$, $\mbol$ and number densities.
It is interesting to note that the predicted dip around $\mj\sim\mk\sim 13$, $\mbol\sim 15$ \cite{Chabrier03} has been confirmed by the most recent L-dwarf
observations \cite{Cruz04}. The dash-line displays the distributions obtained with a power-law IMF
$dn/dm\propto m^{-\alpha}$ with $\alpha=1$, an upper limit for the coefficient \cite{Chabrier02},
with the same normalization at 0.1 $\msol$. As seen in the figure, only in the T dwarf domain do
the two different IMFs yield significantly different densities, a point noted also by Allen et al. (2005).
A more
precise determination of the IMF thus necessitates large statistics in the T dwarf domain. This should be in reach
with future dedicated large field surveys. A census of BDs, stars and stellar remnants in the Galaxy, based on these calculations, and their contribution to the Galactic mass budget
is given in Chabrier (2003) with a slight revision in Chabrier (2005). Extending IMF (1)
down to one jupiter mass yields
BD {\it number} and {\it mass} densities, $n_{BD}\sim n_\star/3\sim 0.03$ pc$^{-3}$ and
$\rho_{BD}\sim \rho_\star/30\sim 1.0\times 10^{-3}\mvol$, respectively. The aforementioned power law IMF with $\alpha=1$ yields at least 5 times more BDs, depending on the minimum mass of the IMF. 

As shown in Chabrier (2003, Figure 10), the IMF (1) yields a mass-to-light ratio in various optical and
infrared bands in very good agreement with dynamical determinations in spiral galaxies \cite{Portinari04},
whereas a Salpeter IMF extending down to the H-burning limit overestimates this ratio by a facto of 2.

\subsection{Young clusters}

As shown in Chabrier (2003, 2005) this field star+BD IMF (1) gives an excellent representation of the IMF of
different young clusters with ages between $\sim 1$ Myr to $\sim 100$ Myrs,
as derived from the observed LFs in various passbands, over the entire observed stellar and brown dwarf domain.
Since these observations do not resolve multiple systems, the IMF in that case is the so-called {\it system} IMF, derived from the field IMF for
resolved objects. The parametrization of this system IMF is given in Chabrier (2003, 2005).
This similarity between field and cluster IMF clearly points towards a rather universal mechanism for star formation in various environments representative of the Galactic disk conditions. This issue is addressed in \S6

\section{Present uncertainties in brown dwarf theory}

As outlined in the previous sections, the present LMS and BD theory has been succesfull in describing, or even in {\it predicting} the observed mechanical and thermal properties of these objects. Several shortcomings, however, remain and there is still room for improvement in various domains.
The description of the spectral energy distribution of cool objects, in particular, needs further improvements. 

\subsection{Young objects ($\sim$ 1-100 Myr)}

Comparison
between observed and sythetic spectra shows an excess (lack) of flux in the J-band (H-band) for the latter ones, whereas
photometry of optical colors shows that the models are too blue by $\sim$ 0.3-0.5 mag \cite{Chabrieretal00a}. Both shortcomings point to the same kind of culprit, namely partially inaccurate
molecular linelists. FeH lists, for example, cover only the 0.9-1.3 $\mu$m range. Water and TiO linelists and/or oscillator strengths
are still imperfect as well. Not only these limitations affect the colors, in particular in the J and H
bands, but they may also bear some impact on the atmospheric structures. Indeed, the derivation of fundamental parameters for young objects, i.e. objects still on the PMS, ranging from $\sim 1$ to $\sim 600$ Myr for low-mass stars, suggest some problems at young ages.
Shortcomings in the theory are indeed more consequential for the evolution of still contracting objects,
characterized by shorter Kelvin-Helmholtz time scales. Song et al. (2002) observed objects near the Lithium burning limit in a 12 Myr cluster
and found no Lithium absorption in objects for which the models
predicted a substantial amount of lithium. White \& Hillenbrandt (2005) reach similar conclusions from their analysis of the optical spectrum
of a spectroscopic binary with an accretion disk in Taurus Auriga.
The age derived from lithium depletion is $\sim 20$ Myr whereas the age inferred from the isochrones {\it at the distance of Taurus}
yields $\sim8 \pm 3$ Myr.
The too low lithium-depletion in the models points to too hot
$\teff$ for a given mass and age (conversely the models suggest too small a mass for a given, observationnally-determined $\teff$).

Caution must be taken, however, before drawing robust conclusions from such an observational analysis. Indeed,
the Song et al. (2002) analysis is based on $\teff$-color relations
characteristic of {\it main sequence} objects \cite{Bessel91} and uncertainties reach {\it at least} $\pm 100$ K.
As for St 34, the object of White \& Hillenbrandt (2005), a subsequent spectroscopic analysis with the Spitzer Telescope \cite{Hartmannetal05}
shows that the object is more likely an older object, possibly as old as 25 Myr, in the foreground of Taurus.
The ages derived from the models using the lithium depletion or the isochrones would then be fully consistent.
Furthermore, all these young objects
show H$_\alpha$ emission lines, a sign of accretion and/or activity. Intense H$_\alpha$ emission implies the presence of a hot ($>10^4$ K), ionizing flux
which may penetrate at least the outer parts of the photosphere, where the cores of Li I resonance lines are formed.
This will yield additional (UV) continuum and overionization and thus reduced equivalent widths of absorption lines.
Indeed, Zapatero-Osorio et al. (2002) and Kenyon et al. (2005) found
significant dispersion in the lithium equivalent widths
of M stars for spectral types cooler than M3.5 in the Sigma Ori cluster. Lithium abundance analysis in the presence of strong H$_\alpha$
emission must thus be considered very cautiously, as what could be considered as lithium depletion might in fact be due
to non-photospheric line formation effect \cite{Pavlenkoetal95}.

In a different analysis, Mohanty et al. (2004) suggest also some inconsistency in the theoretical mass-radius
relationship between the determination based on spectral analysis and the one obtained from evolutionary models. Although, as pointed out by these authors themselves,
gravity and age determinations from the observed spectra remain rather uncertain
for cool objects where the atmosphere is dominated by molecular lines, their careful analysis suggests
too small a gravity (i.e. mass) for a given $\teff$ at a given age, in particular for low-gravity objects. The
same conclusion is obtained from the recent observations of AB Dor C, a low-mass companion to the young ($\sim$ 30-100 Myr) star AB Dor A,
by Close et al. (2005), namely too hot $\teff$, and thus
too bright luminosities from the models for a given mass and age at young ages.
The $\teff$ determination of Close et al. (2005),  however,
is based on an M dwarf $\teff$-Spectral type (Sp) calibration which is not valid for younger, {\it lower gravity}
objects like AB Dor C. Moreover, a new analysis of the age of the AB Dor system \cite{Luhmanetal05a} shows that the
observations are indeed in fairly good (1$\sigma$) agreement with the model predictions.
The same conclusion is reached if AB Dor C is an unresolved BD binary system itself, in which case
the observed $\teff$ and magnitudes would
superbly corroborate the theoretical predictions \cite{Maroisetal05}. Radial velocity analysis or high resolution images should eventually settle this question.

This illustrates how much care must be taken from observational analysis before reaching robust conclusions about the validity of theoretical models. $\teff$ determinations from
spectral types for young (low gravity) objects, in particular, remain very uncertain. Indeed, as shown by Luhman (1999),
young LMS and BD have spectral types which lie between the M dwarf scale and the giant scale and
so far no robust $\teff$-Sp relation exists for objects with in-between gravities. Clearly, the determination of fundamental parameters from observations of young objects remains hampered by substantial
uncertainties and, although theoretical models at young ages certainly need to be improved, conclusions from observational analysis must be considered with due caution.

\subsection{Very young objects ($\la$ 1 Myr)}

Baraffe et al. (2002) have done an extensive analysis of uncertainties in the models for very young ages ($\la$ 1 Myr). As shown by these authors, unknown initial conditions yield different evolutionary paths until at least $\sim 1$ Myr and mass-age determinations from comparison between observations
and evolutionary tracks is totally meaningless in this case.
Other processes, like e.g. the efficiency of convection both in the interior and the atmosphere of low gravity objects remains totaly
undetermined and can modify appreciably the PMS contraction sequence.
Evolution, for example, does not necessarily proceed at constant $\teff$ for a given mass, as is the case only for {\it fully adiabatic} convection in presence of H$_2$ dissociation regions
\cite{Hayashi61, HayashiNakano, BCAH02}.
For this reason,
our group decided not to publish models for such very young ages. It is worrying, however, that
some observers use blindly available models at such ages although at best these models are as inaccurate as
ours (see e.g. Fig. 8 of Baraffe et al. (2002)), and in some cases are based on simplified atmospheric structures and remain inaccurate at any age !
Unfortunately, determinations of young cluster
mass functions based on such erroneous analysis are claimed periodically. We can not stress enough
the necessity to consider such determinations not seriously and to remain aware that
many claimed low-mass IMF determinations for young clusters in the litterature have no reliability !

\section{Brown dwarfs versus planets. Brown dwarf and star formation}

The distinction between BD and giant planets has become these days a topic of intense debate. In 2003,
the IAU has adopted the deuterium-burning minimum mass, $m_{DBMM}\simeq 0.012\,\msol$ (Saumon et al. 1996, Chabrier et al. 2000b)
 as the official distinction between the two types of objects.
We have discussed this limit in previous reviews \cite{Chabrier03, Chabrieretal04} and shown
that it does not rely on any robust physical ground and is a pure semantic definition.
Such an approach is certainly not satisfactory from the scientific point of view and should be abandoned,
for it brings a lot of confusion in the field. This distinction is based on a model of star formation where
deuterium-burning plays a central role (Shu et al. 1977). Indeed deuterium burning generates
a convective instability which in turn leads to the generation of a magnetic wind through dynamo generation. This magneto-centrifugal wind is necessary to sweep away the surrounding accreting gas and thus leads to the formation of a star/BD embryo. This scenario can now be abandoned for various
reasons. First of all observations of star forming regions show that star formation occurs in general on a time scale much shorter than the {\it standard}
ambipolar diffusion time scale characteristic of the previous scenario, namely a few dynamical or turbulent time scales
(see e.g. Hartmann et al. (2001), Kirk et al. (2005) their Fig. 11).
Second, all BD interiors are convective, with or without deuterium burning. Although, admitedly, this holds
for a non-accreting, or at least non spherically accreting protostar and has not been demonstrated to
remain true for an accreting object. The phase of spherical accretion, the so-called "class 0" phase, however, is observed to take place over
a very short time scale \cite{Andreetal00}. Very rapidly accretion is taking place through a disk or along magnetic poles
\cite{Bontempsetal96, Henriksenetal97} and thus the internal structure
of the protostar very likely remains convective \cite{Hartmannetal97}. Finally, even though magnetic winds may play a role in star/BD
formation, it is unlikely that they play a dominant role and gas reservoirs
limited by turbulence are appealing alternative mechanisms to limit the accretion on the central core. In any event,
as mentioned above, deuterium burning can {\it not} be considered as a necessary mechanism to trigger star formation and thus
can not be considered as a robust limit for star formation.
In fact the observation of free floating objects with masses of the order of a few jupiter masses in (low extinction) young clusters
\cite{Bejaretal01} shows that star and BD formation extends down to jupiter-like masses.

The $\sim$10 jupiter-mass limit, however, illustrates also the so-called opacity-limited minimum mass for fragmentation \cite{LowLyndenBell76, Silk77}.
This limit is generally considered as robust because of the weak dependence of this minimum mass upon (dust) opacity, $m_{min}\propto {\bar \kappa}^{1/7}$.
Effects like magnetic field \cite{Boss01}, anisotropy or shocks \cite{BoydWhitworth05}, however, can significantly lower this limit, down to a few jupiter masses.

We thus suggest the following clarification for "star", "brown dwarf", "sub brown dwarf", "planetary object" or "planet" denominations.
Objects formed either in systems or as isolated objects from the collapse of a cloud are either stars or brown dwarfs.
The {\it physically grounded} limit between the two types of objects is the hydrogen-burning minimum mass, $m\sim 0.07\,\msol$
for solar abundance \cite{CB97},
below which the object {\it can not reach thermal equilibrium} and keeps contracting steadily with time (Chabrier \& Baraffe 2000 Fig. 2).
Given this, there are deuterium-burning brown dwarfs and {\it non} deuterium burning brown dwarfs, depending whether their mass exceeds or not the
deuterium-burning minimum mass, exactly as stars above $\sim 1.5\,\msol$ ignite the CNO burning cycle in their core and stars below this
mass do not. Both types of objects, however, are called stars and not "substars" or other exotic name ! Planets, on the other hand, are objects
formed in a protoplanetary disk around a parent star, with a high mass ratio. They should thus have enhanced average abundances of heavy elements.
Although the {\it direct} spectroscopic observational confirmation of such a diagnostic remains for now out
of reach, the recent observation of the transiting planet HD 149026B, with an inferred 70 M$_\oplus$
of heavy element for a planet mass $m\,\sin i=0.36\,$ m$_{jup}$ (Sato et al. 2005) seems to comfort this suggestion.
Therefore, the various "direct" detections of exoplanets at large orbital distances ($\ga 50$ AU) claimed recently
in the literature, and based on the aforementioned IAU definition, are most likely in reality wide binary {\it brown dwarfs} (see Gizis et al. (2005)),
 by itself an interesting result.

\begin{figure}[!ht]
\plotone{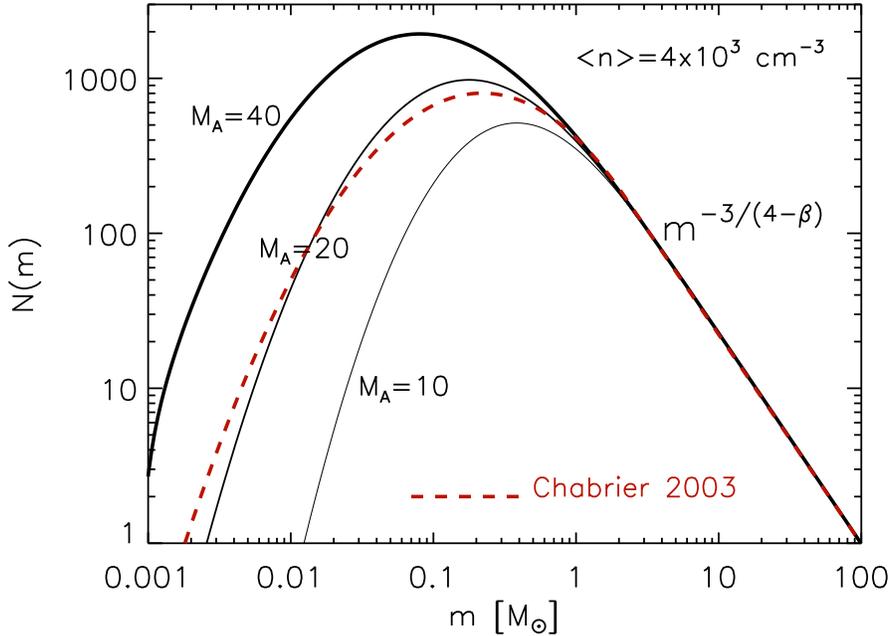}
%\vspace{2.8in}
\caption{Comparison between the {\it system} IMF representative of the Galactic disk and young clusters (Chabrier 2003)
and the distribution of core masses obtained from compressible MHD turbulence (Padoan \& Nordlund 2002) }
\label{FigMHD}
\end{figure}

This brings us to the issue of star and BD formation. As mentioned above, the conventional scenario for star formation, which assumes
that proto-stellar cores are in quasi-static equilibrium against gravitation and form stars once magnetic support is lost
through ambipolar diffusion \cite{Shuetal77},
can now be abandoned on various grounds. The modern picture of star formation suggests that cores are formed by compressible motions in the
turbulent velocity field of the cloud
\cite{BallesterosParedesetal03, Klessenetal05, PadoanNordlund02}. Indeed, large scale ($\ge 0.1$ pc) cloud structures
are dominated by (supersonic) turbulence. At small (stellar) scales,
gravity takes over and fragmentation becomes gravitational rather than turbulent, with only the densest cores with gravitational energy
exceeding the local internal energy collapsing into stars. In this scenario, progressively smaller density peaks contain progressively smaller
mass fractions, yielding decreasing star formation efficiency with decreasing scales below the Jeans mass \cite{PadoanNordlund02, VazquezSemadenietal03}.
In these calculations, BDs represent simply the low-mass part of the star formation process \cite{PadoanNordlund04}.
Figure \ref{FigMHD} compares the mass spectrum obtained
from MHD simulations of supersonic, superAlf\'venic turbulence \cite{PadoanNordlund02} with the {\it system} IMF obtained for the Galactic disk and young clusters in \S4
 (indeed, the identified prestellar cores in the simulation correspond to multiple systems, for some of these cores might eventually become unstable and fragment into more individual
objects). Although such a comparison must be considered with
caution before drawing any conclusion (e.g. gravity, which will be dominant at the stellar scale, is not included in the simulations
yet), the at least qualitative and even semi-quantitative agreement between the simulations and the
IMF representative of the field and young clusters is encouraging. This shows that BDs form {\it in adequat numbers}
from the same fragmentation mechanism as for star formation. Various
observations of disk accreting BDs show indeed that BDs and stars form from the same underlying mechanism
\cite{WB, Jaya03, Luhmanetal05b, Gizis05, Kenyonetal05}.
Note that in these simulations of supersonic turbulence,
only a few percents of the total mass end up into the collapsing cores
after one dynamical time, providing a solution to
the problematic high efficiency problem associated with turbulence-driven star formation. Although admitedly time scale may remain an
issue.
This new picture thus combines turbulence, as the initial driving mechanism for fragmentation, and gravity, providing a natural explanation for a (scale free)
power-law IMF above a critical mass, namely the mean thermal Jeans mass $\langle m_J \rangle$, and a lognormal distribution below, due to the fact that only the densest cores, exceeding the local Jeans mass, will collapse into bound objects.

The alternative scenario of formation of BDs by ejection \cite{ReipurthClarke01, Boss01, B02} can now be abandoned as a {\it dominant} scenario
for BD formation for various reasons. First of all, the aforementioned observations of disk accretion in BDs exhibit characteristics similar to the
more massive T Tauri stars. Second of all, the recent observations of wide binary BDs \cite{Luhman04, Forveilleetal04, Chauvinetal05, Billieresetal05} contradict the predictions
of such a scenario as a dominant formation mechanism for BDs. At last, the observed high binary frequency of VLMS/BDs,
in particular for close binaries, rules out at
$>$99 confidence level a dynamical origin for VLMS/BDs, as obtained from N-body models (which neglect in particular gas-dynamic hardening mechanisms) or
hydrodynamical SPH simulations \cite{MaxtedJeffries05}.

\section{Conclusion and perspectives}

As mentioned in the introduction, accurate models for low-mass stars, brown dwarfs and giant
planets
are needed to shed light on the observable properties of these objects and to
provide guidance to the ongoing and future surveys aimed at understanding their formation, structure and evolution and
at revealing their contribution
to the Galactic population. Tremendous progress has been realyzed from this point of view
within the recent years. As outlined in this review, the present modern theory of LMS/BD structure and evolution
has been very succesfull in accurately predicting several stringent observational constraints : spectra,
mass-magnitude and mass-radius relationships, color-magnitude diagrams.  
Any theory aimed at describing the mechanical and thermal properties of these objects {\it must} be confronted to these experimental/observational constraints in order to assess its degree of validity.
Improvement in the theory is still necessary, in particular for young, low gravity objects, and will proceed along with the discovery of many more substellar objects, down to
Jupiter-like masses.

Present determinations of the star+BD IMF give satisfactory comparisons with observational determinations and yield reasonably robust number-density and mass-density estimates for the stellar
and BD census in the Galaxy.
The increasing number of observed BDs will improve the accuracy of these determinations.
Although such an IMF can be described by a series of power laws, such an approach, though
convenient, is not satisfactory form the physics point of view as it can not be rooted to any physical
mechanism for star formation. Indeed any change in the slope implies a different characteristic mass.
A general Salpeter-like power law above the average thermal Jeans mass ($\sim 1\msol$ for present
day conditions) rolling down into a lognormal form below this mass, on the contrary, is consistent with
physical mechanisms like turbulence-driven fragmentation. Although a complete theory for star/BD formation
is still elusive, turbulence fragmentation at large scale indeed provides an appealing solution for
the "universal" triggering mechanism and leads to the formation of stars and BDs from the same
initial process. Processes like magnetic winds, accretion, dynamical interactions
will certainly play some role in shaping up the final IMF but they do not seem to play a {\it dominant} role
in the underlying general process.

At last, we argue that the deuterium-burning limit as a definition of BD vs planet distinction should be abandoned,
for it does not fit into a modern approach of star formation, and BD formation very likely extends below
this limit.

{}


\begin{thebibliography}{}
\bibitem[Ackerman \& Marley 2001]{AckermanMarley01} Ackerman, A., \& Marley, M.,  2001, \apj, 556, 872
\bibitem[Allard et al. 1996]{Allardetal96} Allard, F., Hauschildt, P.H., Baraffe, I., \& Chabrier, G., 1996,
\apjl, 465, L123
\bibitem[Allard et al. 2001]{Allardetal01} Allard F., Hauschildt P.H., Alexander D.R., Tamanai A.,
Ferguson J., 2001, \apj, 556, 357
\bibitem[Allard et al. 2003b]{AllardAllard} Allard, N., Allard F., Hauschildt, P., Kielkopf, J., \& Machin, L., 2003, \aap, 411, L473
\bibitem[Allard et al. 2003a]{Allardetal03} Allard, F., et al., in {\it Brown Dwarfs}, IAU Symposium \#211, Astronomical Society of the Pacific, p. 325
\bibitem[Allen et al. 2005]{Allen05} Allen, P., Koerner, D., Reid, N., \& Trilling, D.,  2005, \apj, 625, 385
\bibitem[Andersen 1991]{Andersen91} Andersen, J., 1991, \araa, 3, 91
\bibitem[Andr{\'e} et al. 2000]{Andreetal00} Andr{\'e}, P., Ward-Thomson, D., \& Barsony, M., 2000,
in {\it Protostars and Planets IV}, University of Arizona Press, p. 59
\bibitem[Ballesteros-Paredes et al. 2003]{BallesterosParedesetal03} Ballesteros-Paredes, J., Klessen, R., V\'azquez-Semadeni, E., 2003, \apj, 592, 188
\bibitem[Baraffe at al. 1998]{BCAH98} Baraffe, I., Chabrier, G., Allard, F. and Hauschildt, P.H., 1998, \aap, 337, 403
\bibitem[Baraffe et al. 2002]{BCAH02} Baraffe, I., Chabrier, G., Allard, F. and Hauschildt, P.H., 2002, \aap, 382, 563
\bibitem[Baraffe et al. 2003]{Baraffeetal03} Baraffe, I., Chabrier, G., Barman, T., Allard, F. and Hauschildt, P.H., 2003, \aap, 402, 701
\bibitem[Bate et al. 2002]{B02} Bate, M., Bonnell, I., \& Bromm, V., 2002, \mnras, 332, L65
\bibitem[Borysow et al. 1985]{Borysowetal} Borysow, A., Trafton, L., Frommhold, L. \& Birnbaum, G., 1985, \apj, 296, 644
\bibitem[Bejar et al. 2001]{Bejaretal01} Bejar, V., et al., 2001, \apj, 556, 830
\bibitem[Belov et al. 2002]{Belovetal02} Belov et al., 2002, Soviet-Phys-JETP Lett, 76, 433
\bibitem[Bessel 1991]{Bessel91} Bessel, M., 1991, \aj, 101, 662
\bibitem[Billi\`eres et al. 2005]{Billieresetal05} Billi\`eres, M., et al., 2005, \aap, in press
\bibitem[Bontemps et al. 1996]{Bontempsetal96} Bontemps, S., Andr{\'e}, P., Terebey, S., \& Cabrit, S., 1996, \aap, 311, 858
\bibitem[Boriskov et al. 2003]{Boriskovetal03} Boriskov et al, 2003, Dokl. Phys., 48, 553
\bibitem[Boss 2001]{Boss01} Boss, A., 2001, \apjl, 551, L167
\bibitem[Boyd \& Whitworth 2005]{BoydWhitworth05} Boyd, D., \& Whitworth, A.,  2005, \aap, 430, 1059
%\bibitem[Burgasser et al. 2000]{Burgasseretal00} Burgasser et al. 2000
\bibitem[Burgasser 2001]{Burgasser01} Burgasser, A., 2001, PhD Thesis, California Institute of Technology
\bibitem[Burgasser et al. 2002]{Burgasseretal02} Burgasser, A., et al. 2002, \apj, 571, L151
\bibitem[Burgasser 2004]{Burgasser04} Burgasser, A., 2004, \apjs, 155, 191
\bibitem[Burrows \& Sharp 1999]{BurrowsSharp99} Burrows A., \& Sharp, J., 1999, \apj, 512, 843
\bibitem[Burrows et al. 2000]{Burrowsetal00} Burrows A., Marley, M.,  \& Sharp, J., 2000, \apj, 531, 438
\bibitem[Burrows \& Volobuyev 2003]{BurrowsX} Burrows A., \& Volobuyev, 2003, \apj, 583, 985
\bibitem[Chabrier 2002]{Chabrier02} Chabrier, G., 2002, \apj, 567, 304
\bibitem[Chabrier 2003]{Chabrier03} Chabrier, G., 2003, \pasp, 115, 763
\bibitem[Chabrier 2005]{Chabrier05} Chabrier, G., 2005, in {\it The initial mass function 50 years later}, Springer, 327, 41
\bibitem[Chabrier \& Baraffe 1997]{CB97} Chabrier, G., \& Baraffe, I., 1997, \aap, 327, 1039
\bibitem[Chabrier \& Baraffe 2000]{CB00} Chabrier, G., \& Baraffe, I., 2000, \araa, 38, 337
\bibitem[Chabrier et al. 2000a]{Chabrieretal00a} Chabrier, G., Baraffe, I., Allard, F. and Hauschildt, P.H., 2000a, \apj, 542, 464
\bibitem[Chabrier et al. 2000b]{Chabrieretal00b} Chabrier, G., Baraffe, I., Allard, F. and Hauschildt, P.H., 2000b, \apj, 542, L119
\bibitem[Chabrier et al. 2004]{Chabrieretal04} Chabrier, G., Baraffe, I., Allard, F. and Hauschildt, P.H., 2004, in {\it Extrasolar Planets: Today and Tomorrow}, ASPC, 321, 131
\bibitem[Chauvin et al. 2005]{Chauvinetal05} Chauvin, G., et al., 2005, \aap, 438, L25
\bibitem[Close et al. 2005]{Closeetal05} Close, L., et al. 2005, Nature, 433, 286
\bibitem[Collins et al. 1998]{Collinsetal98}  Collins, G.W., \etal, 1998, {\it Science}, 281, 1178
\bibitem[Cruz 2004]{Cruz04} Cruz.., K-L., 2004, PhD Thesis, University of Pennsylvania
\bibitem[Dahn et al. 1986]{Dahnetal86} Dahn C.C., Libert, J,m \& Harrington, R., 1986, \aj, 91, 621
\bibitem[Dahn et al. 2002]{Dahnetal02} Dahn C.C., et al., 2002, \aj, 124, 1170
\bibitem[Delfosse et al. 2000]{Del00} Delfosse, X., et al., 2000, \aap, 364, 217
\bibitem[Fegley \& Lodders 1996]{FegleyLodders96} Fegley B., \& Lodders K., 1996, \apj, 472, L37
\bibitem[Forveille et al. 2004]{Forveilleetal04} Forveille, T., et al., 2004, \aap, 427, L1
\bibitem[Gizis et al. 2000]{Gizis00} Gizis, J., et al., 2000, \aj, 120, 1085
\bibitem[Gizis et al. 2005]{Gizis05} Gizis, J.,  Shipman, H., \& Harvin, J., 2005, astro-ph/0507429
\bibitem[Hayashi 1961]{Hayashi61} Hayashi, C., PASJ, 1961, 13, 450
\bibitem[Hayashi \& Nakano 1963] {HayashiNakano} Hayashi, C., \& Nakano, 1963, Progr. Theor. Physics, 30, 460
\bibitem[Hartmann et al. 1997]{Hartmannetal97} Hartman L., Cassen, P., Kenyon, S.J., 1997, \apj, 475, 770
\bibitem[Hartmann et al. 2001]{Hartmannetal01} Hartmann, L., Ballesteros-Paredes, J., \& Bergin, E., 2001, \apj, 562, 852
\bibitem[Hartmann et al. 2005]{Hartmannetal05} Hartmann, L., et al., 2005, \apj, 628, L147
\bibitem[Hauschildt et al. 1999]{Hauschildtetal99} Hauschildt, P.H., Allard, F., Baron, E., 1999, \apj, 512, 377
\bibitem[Henriksen et al. 1997]{Henriksenetal97} Henriksen, R., Andr\'e, P., Bontemps, S., 1997, \aap, 323, 549
\bibitem[Henry \& McCarthy 1990]{HmcC90} Henry, T., \& McCarthy, D., 1990, \apj, 350, 334
\bibitem[Jayawardhana et al. 2003]{Jaya03} Jayawardhana, R., Mohanty, S., \& Basri, G., 2003, \apj, 592, 282
\bibitem[Jones \& Tsuji 1997]{JonesTsuji97} Jones, H., \& Tsuji, T., 1997, \apj. 480, L39
\bibitem[Kenyon et al. 2005]{Kenyonetal05} Kenyon, M., Jeffries, R., Naylor, T., Oliveira, J., \& Maxted, P., 2005, \mnras, 356, 89
\bibitem[Kirk et al. 2005]{Kirketal05} Kirk, J., Ward-Thomson, D., \& Andr\'e, P., 2005, \mnras, 360, 1506
\bibitem[Kirkpatrick et al. 1999]{Kirkpatricketal99} Kirkpatrick, J. D., Allard, F., Bida, T., Zuckerman, B., Becklin, E.E., Chabrier, G., and Baraffe, I., 1999, \apj, 519, 834
\bibitem[Kirkpatrick et al. 2000]{Kirkpatricketal00} Kirkpatrick, J. D.,  et al., 2000, \aj, 120, 473
\bibitem[Klessen et al. 2005]{Klessenetal05} Klessen, R., Ballesteros-Paredes, V\'azquez-Semadeni, \& Dur\'an-Rojas, C., 2005, \apj, 620, 786
\bibitem[Knapp et al. 2004]{Knappetal04} Knapp, G., et al., 2004, \aj, 127, 3553
\bibitem[Knudson et al. 2004]{Knudsonetal04} Knudson, M., et al., 2004, \prb, 69, 144209
\bibitem[Kroupa 2001]{KTG} Kroupa, P, 2001, \mnras, 322, 231
\bibitem[Kroupa et al. 1993]{KTG} Kroupa, P, Tout, C, \& Gilmore, G, 1993, \mnras, 262, 545
\bibitem[Leggett et al. 1998]{Leggettetal98} Leggett S.K., Allard, F., \& Hauschildt, P.H., 1998, \apj, 509, 836
\bibitem[Linsky 1969]{Linsky69} Linsky J. 1969. \apj. 156, 989
\bibitem[Lodders 1999]{Lodders99} Lodders, K., 1999, \apj, 519, 793
\bibitem[Low \& Lynden-Bell 1976]{LowLyndenBell76} Low, C., \& Lynden-Bell, D., 1976, \mnras, 176, 367
\bibitem[Luhman 1999]{Luhman99} Luhman, K., 1999, \apj, 525, 466
\bibitem[Luhman et al. 2003]{Luhmanetal03} Luhman, K., et al., 2003, \apj, 593, 1093 (see Appendix B)
\bibitem[Luhman 2004]{Luhman04} Luhman, K., 2004, \apj, 614, L398
\bibitem[Luhman et al. 2005a]{Luhmanetal05a} Luhman, K., Stauffer, J., \& Mamajek, E., 2005, \apjl, 628, L69
\bibitem[Luhman et al. 2005b]{Luhmanetal05b} Luhman, K., et al., 2005, \apjl, 620, L51
\bibitem[Marley et al. 1996]{Marleyetal96} Marley, M.S., Saumon, D., Guillot, T., Freedman, R., Hubbard, W.B., Burrows, A., Lunine, J.I. 1996, Science, 272, 1919
\bibitem[Marley et al. 2002]{Marleyetal02} Marley, M.S., et al., 2002, \apj, 568, 335
\bibitem[Maxted \& Jeffries 2005]{MaxtedJeffries05} Maxted, P., \& Jeffries, R., 2005, \mnras, in press
\bibitem[Marois et al. 2005]{Maroisetal05} Marois, C., Macintosh, B., Song, I., \& Barman, T., 2005, astro-ph/0502382
\bibitem[Nellis et al. 1983]{Nellis83} Nellis, W.,  et al., 1983, J. Chem. Phys., 79, 1480
\bibitem[Mohanty et al. 2004]{Mohantyetal04}   Mohanty, S., Jayawardhana, R., \& Basri, G., 2004, \apj, 609, 885
\bibitem[Noll et al. 1997]{NollGeballeMarley97} Noll, K., Geballe, T., \& Marley, M., 1997, \apj, 489, L87
\bibitem[Oppenheimer et al. 1995]{Oppenheimeretal95} Oppenheimer, B.R., Kulkarni, S.R., Nakajima, T., Matthews, K. 1995, Science, 270, 1478
\bibitem[Padoan \& Nordlund 2002]{PadoanNordlund02} Padoan \& Nordlund 2002, \apj, 576, 870
\bibitem[Padoan \& Nordlund 2004]{PadoanNordlund04} Padoan \& Nordlund 2004, \apj, 617, 559
\bibitem[Pavlenko et al. 1995]{Pavlenkoetal95} Pavlenko, Y.V., Rebolo, R., Mart\'{i}n, E., \& Garc\'{i}a L\'{o}pez, R.J., 1995, \aap, 303, 807
\bibitem[Pont et al. 2005]{Pontetal05} Pont, F., et al., 2005, \aap, 433, L21
\bibitem[Portinari et al. 2004]{Portinari04} Portinari, L., Sommer-Larsen, J., \& Tantalo, R., 2004, \mnras, 347, 691
\bibitem[Reid et al. 1999]{Reidetal99} Reid I.N., et al., 1999, \apj. 521, 613
\bibitem[Reid et al. 2002]{Reid02}  Reid, N., Gizis, J., \& Hawley, S., 2002, \aj, 124, 2721
\bibitem[Reipurth \& Clarke 2001]{ReipurthClarke01} Reipurth, B., Clarke, C., 2001, \aj, 122, 432
\bibitem[Ruiz et al. 1997]{Ruizetal97} Ruiz M., Leggett S.K., Allard F., 1997, \apj, 491, L107
\bibitem[Sato et al. 2005]{Satoetal05} Sato, B., et al., 2005, \apj, in press, astro-ph/0507009
\bibitem[Saumon, Chabrier \& VanHorn 1995]{SCVH95} Saumon, D., Chabrier, G., and VanHorn, H.M., 1995, \apjs, 99, 713
\bibitem[Saumon et al. 2000]{Saumonetal00} Saumon, D., Chabrier, G., Xu, \& Wagner, 2000, High Pressure Research, 16, 331
\bibitem[Saumon et al. 1996]{Saumonetal96} Saumon, D, Hubbard, W., Burrows, A., Guillot, T., Lunine, J., Chabrier, G., 1996. \apj. 460, 993
\bibitem[Saumon et al. 2003]{Saumonetal03} Saumon D, Marley, M., Lodders, K., \& Freedman, R., 2003, in {\it Brown Dwarfs}, IAU Symposium \#211, Astronomical Society of the Pacific, p. 345
\bibitem[Segransan et al. 2003]{Seg03} Segransan, D., Kervella, P., Forveille, T., \& Queloz, D., 2003, \aap, 397, L5
\bibitem[Segransan et al. 2003]{Seg03b} Ségransan, D., Delfosse, X., Forveille, T., Beuzit, J. L., Perrier, C., Udry, S., Mayor, M.,
 in {\it Brown Dwarfs}, IAU Symposium \#211, Astronomical Society of the Pacific, p. 413
\bibitem[Shu et al. 1977]{Shuetal77} Shu, F., Adams, F., \& Lizano, S., 1987, \araa, 25, 23
\bibitem[Silk 1977]{Silk77} Silk, J., \apj, 214, 152
\bibitem[Song et al. 2002]{Songetal02} Song, I., Bessell, M., \& Zuckerman, B., 2002, \apj, 581, L43
\bibitem[Tinney et al. 1998]{Tinneyetal98} Tinney, C., Delfosse, X., Forveille, T., \& Allard, F., 1998, \aap, 338, 1066
\bibitem[Tinney et al. 2003]{TinneyBurgasserKirkpatrick03} Tinney, C., Burgasser, A., \& Kirkpatrick, J., 2003, \aj, 126, 975
\bibitem[Torres \& Ribas 2002]{TorresRibas02} Torres, G., \& Ribas, I., 2002, \apj, 567, 1140
\bibitem[Tsuji et al. 1996]{Tsujietal96} Tsuji, T., Ohnaka, K. and Aoki, W. 1996, \aap, 305, L1
\bibitem[Tsuji et al. 1999]{Tsujietal99} Tsuji, T., Ohnaka, K. and Aoki, W. 1996, \apj, 520, L119
\bibitem[Tsuji 2005]{Tsuji05} Tsuji, T., 2005, \apj, 621, 1033
\bibitem[V\'azquez-Semadeni et al. 2003]{VazquezSemadenietal03}  V\'azquez-Semadeni, E., Ballesteros-Paredes, \& J., Klessen, R., 2003, \apj, 585, L131
\bibitem[Vrba et al. 2004]{Vrbaetal04} Vrba, F., et al., 2004, \aj, 127, 2948
\bibitem[White et al. 1999]{Whiteetal99} White, R., Ghez, A., Reid, I., Schultz, G., 1999,  \apj, 520, 81
\bibitem[White \& Hillenbrandt 2005]{WhiteHillenbrandt} White, R., \& Hillenbrandt, L., 2005, \apj, 621, L65
\bibitem[White \& Basri 2003]{WB} White, R., \& Basri, G., 2003, \apj, 582, 1109
\bibitem[Zapatero-Osorio et al. 2002]{ZapateroOsorioetal02} Zapatero-Osorio, M.R., et al., 2002, \aap, 384, 937
\bibitem[Zoccali et al. 2000]{Zoccali00} Zoccali, M., et al., 2000, \apj, 530, 418

\end{thebibliography}
\end{document}